\documentclass[12pt,a4paper]{article}

\usepackage[left=2.5cm,top=2.5cm,bottom=2.5cm,right=2.5cm]{geometry}
\usepackage{graphicx}
\usepackage{amssymb}
\usepackage{amsthm}
\usepackage{amsmath}
\usepackage{amsfonts}
\usepackage{algorithm, algorithmic}
\usepackage{color}

\usepackage{cite}

\usepackage{enumitem}

\newtheorem{thm}{Theorem}[section]

\newtheorem{lem}[thm]{Lemma}

\theoremstyle{definition}

\theoremstyle{remark}
\newtheorem{rem}{Remark}[section]

\numberwithin{equation}{section}

\DeclareMathOperator{\idiv}{\,{\bf div } \,}
\DeclareMathOperator{\imod}{\,{\bf mod } \,}

\begin{document}

\title{Sorting distinct integers using improved in-place associative sort}


\author{A. Emre CETIN \\
email: aemre.cetin@gmail.com}

\maketitle

\begin{abstract}

{\em In-place associative integer sorting} technique was proposed for integer lists which requires only constant amount of additional memory replacing bucket sort, distribution counting sort and address calculation sort family of algorithms. Afterwards, the technique was further improved and an in-place sorting algorithm is proposed where $n$ integers $S[0 \ldots n-1]$ each in the range $[0, n-1]$ are sorted exactly in $\mathcal{O}(n)$ time while the complexity of the former technique was the recursion $T(n) = T(\frac{n}{2}) + \mathcal{O}(n)$ yielding $T(n) = \mathcal{O}(n)$.

The technique was specialized with two variants one for read-only distinct integer keys and the other for modifiable distinct integers, as well. Assuming $w$ is the fixed word length, the variant for modifiable distinct integers was capable of sorting $n$ distinct integers $S[0 \ldots n-1]$ each in the range $[0, m - 1]$ in exactly $\mathcal{O}(n)$ time if $m < (w-\log n) n$. Otherwise, it sort in $\mathcal{O}(n+\frac{m}{w-\log n})$ time for the worst,  $\mathcal{O}(\frac{m}{w-\log n})$ time for the average (uniformly distributed keys) and $\mathcal{O}(n)$ time for the best case using only $\mathcal{O}(1)$ extra space. 

In this study, the variant for modifiable distinct integers is improved and an algorithm is obtained that sorts $n$ distinct integers $S[0 \ldots n-1]$ each in the range $[0, m - 1]$ in exactly $\mathcal{O}(n)$ time if $m < (w-1) n$. Otherwise, it sort in $\mathcal{O}(n+\frac{m}{w-1})$ time for the worst,  $\mathcal{O}(\frac{m}{w-1})$ time for the average (uniformly distributed keys) and $\mathcal{O}(n)$ time for the best case using only $\mathcal{O}(1)$ extra space.

\end{abstract}


\section{Introduction}\label{sec:intro}

Nervous system is considered to be closely related and described with the ``serial order in behavior" in cognitive neuroscience~\cite{Lashley,Lashley_1} with three basic theories which cover almost all {\em abstract data types} used in computer science. These are~\cite{Henson} chaining theory, positional theory and ordinal theory.

Chaining theory is the extension of stimulus-response (reflex chain) theory, where each response can become the stimulus for the next. From an information processing perspective, comparison based sorting algorithms that sort the lists by making a series of decisions relying on comparing keys can be classified under chaining theory. Each comparison becomes the stimulus for the next. Hence, keys themselves are associated with each other. Some important examples are quick sort~\cite{Hoare}, shell sort~\cite{Shell}, merge sort~\cite{Burnetas} and heap sort~\cite{Williams}.

Positional theory assumes order is stored by associating each element with its position in the sequence. The order is retrieved by using each position to cue its associated element. Content-based sorting algorithms where decisions rely on the contents of the keys can be classified under this theory. Each key is associated with a position depending on its content. Some important examples are distribution counting sort~\cite{Seward,Feurzig}, address calculation sort~\cite{Isaac,Tarter,Flores,Jones,Gupta,Suraweera}, bucket sort\cite{mahmoud:2000, Cormen} and radix sort~\cite{knuth:vol3,mahmoud:2000,sedgewick:algorithms_in_C, Cormen}.

Ordinal theory assumes order is stored along a single dimension, where that order is defined by relative rather than absolute values on that dimension. Order can be retrieved by moving along the dimension in one or the other direction. This theory need not assume either the item-item nor position-item associations of the previous theories.

The main difficulties of all distributive sorting algorithms is that, when the keys are distributed using a hash function according to their content, several keys may be clustered around a loci, and several may be mapped to the same location. These problems are solved by inherent three basic steps of associative sort~\cite{ecetin} (i) {\em practicing}, (ii) {\em storing} and (iii) {\em retrieval} which are the three main stages in the formation and retrieval of memory in cognitive neuroscience.

\section{Original Technique}

As in the ordinal model of Shiffrin and Cook\cite{Shiffrin,Henson}, it is assumed that associations are between the integers in the list space and the nodes in an imaginary linear subspace (ILS) that spans a predefined range of integers. The ILS can be defined anywhere on the list space $S[0\ldots n-1]$ provided that its boundaries do not cross over that of the list. The range of the integers spanned by the ILS is upper bounded by the number of integers $n$ but may be smaller and can be located anywhere making the technique in-place, i.e., beside the input list, only a constant amount of memory locations are used for storing counters and indices. An association between an integer and the ILS is created by a node using a monotone bijective hash function that maps the integers in the predefined interval to the ILS. The process of creating a node by mapping a distinct integer to the ILS is ``practicing a distinct integer of an interval''. Once a node is created, the redundancy due to the association between the integer and the position of the node releases the word allocated to the integer in the physical memory except for one bit which tags the word as a node for interrogation purposes. The tag bit discriminates the word as node and the position of the node lets the integer be retrieved back from the ILS using the inverse hash function. This is ``integer retrieval". All the bits of the node except the tag bit can be cleared and used to encode any information. Hence, they are the ``record'' of the node and the information encoded into a record is the ``cue'' by which cognitive neuro-scientists describe the way that the brain recalls the successive items in an order during retrieval. For instance, it will be foreknown from the tag bit that a node has already been created while another occurrence of that particular integer is being practiced providing the opportunity to count other occurrences. The process of counting other occurrences of a particular integer is ``practicing all the integers of an interval'', i.e., rehearsing used by cognitive neuro-scientists to describe the way that the brain manipulates the sequence before storing in a short (or long) term memory. Practicing does not need to alter the value of other occurrences. Only the first occurrence is altered while being practiced from where a node is created. All other occurrences of that particular integer remain in the list space but become meaningless. Hence they are ``idle integers''. On the other hand, practicing does not need to alter the position of idle integers as well, unless another distinct integer creates a node exactly at the position of an idle integer while being practiced. In such a case, the idle integer is moved to the former position of the integer that creates the new node. This makes associative sort unstable, i.e., equal integers may not retain their original relative order. 

Once all the integers in the predefined interval are practiced, the nodes dispersed in the ILS are clustered in a systematic way closing the distance between them to a direction retaining their relative order. This is the {\em storing} phase of associative sort where the received, processed and combined information to construct the sorted permutation of the practiced interval is stored in the short-term memory. When the nodes are moved towards a direction, it is not possible to retain the association between the ILS and list space. However, the record of a node can be further used to encode the absolute (former) position of that node as well, or maybe the relative position or how much that node is moved relative to its absolute or relative position during storing. Unfortunately, this requires that a record is enough to store both the positional information and the number of idle integers practiced by that node. However, as explained earlier, further associations can be created using the idle integers that were already practiced by manipulating either their position or value or both. Hence, if the record is enough, it can store both the positional information and the number of idle integers. If not, an idle integer can be associated accompanying the node to supply additional space for it for the positional information.

Finally, the sorted permutation of the practiced interval is constructed in the list space, using the stored information in the short-term memory. This is the {\em retrieval} phase of associative sort that depends on the information encoded into the record of a node. If the record is enough, it stores both the position of the node and the number of idle integers. If not, an associated idle integer accompanying the node stores the position of the node while the record holds the number of idle integers. The positional information cues the recall of the integer using the inverse hash function. This is ``integer retrieval'' from imaginary subpace. Hence, the retrieved integer can be copied on the list space as many as it occurrs.

Hence, moving through nodes that represent the start and end of practiced integers as well as retaining their relative associations with each other even when their positions are altered by cuing allow the order of integers to be constructed in linear time in-place.

\subsection{Improved Technique}

With a simple revision~\cite{ecetin2}, the associative sorting technique is improved both theoretically and practically  and a faster technique is achieved. During storing where the nodes are clustered at the beginning of the list retaining their relative order, the positional information ($\log n$ bits) of a node is encoded into either its record or an idle-integer accompanying the node. However, the tag bit discriminates the word as a node in the list space and if ignored during storing it will continue to discriminate the word as a node. This means that, if only the records ($w-1$ bits) of the nodes are clustered at the beginning of the list (short-term memory) retaining their relative order, there will be $n_d$ nodes dispersed in the list space, and $n_d$ records in the short-term memory ($S[0 \ldots n_d-1]$) after storing. Hence, a one-to-one correspondence is obtained with the clustered records and the nodes (tagged words) of the list. Therefore, retrieval phase can search the list from right to left for the first tagged word, retrieve the integer from the ILS through that node, read its number of occurrence from its record $S[n_d-1]$ in the short-term memory and expand it over the list starting at $S[n_d+n_c-1]$ where $n_c$ is the number of practiced idle integers. Afterwards, the processed tag bit can be cleared and a new search to the left can be carried for the next tagged word which will correspond to the next record $S[n_d-2]$ of the short-term memory. This can continue until all the integers are retrieved from short-term memory resulting in the sorted permutation of the practiced integers.

\subsection{Sorting Distinct Integers with Original Technique}

If it is known that all the integers of the list are distinct, associative sorting technique can be specialized~\cite{ecetin1} because there is only one integer that can be practiced and mapped to a location creating a node. Two solutions are possible in this case. The first one is for read-only keys and instead of tagging the word as node using its most significant bit (MSB), the key itself can be used to tag the word ``implicitly'' as node without modifying it, since when a key is mapped to the ILS, it will always satisfy the monotone bijective hash function. The keys are ``implicitly practiced'' in this case. Hence, storing phase is enough to obtain the sorted permutation of the practiced interval cancelling the retrieval phase. In each iteration only the keys that fall into the range $[\delta, \delta+n-1]$ can be sorted where $\delta$ is the minimum of the list of that iteration. It should be noted that, this variant is suitable for sorting a list $S$ of $n$ {\em elements}, $S[0 \ldots n-1]$ each have an integer {\em key} where the problem is to sort the elements of the list according to their integer keys. 

The other scenario is that, when a distinct integer is mapped to the ILS, its record can be used to improve the interval of range of integers to be practiced. During storing, each node is clustered at the beginning of the list together with its record retaining its relative order with respect to others. At this point, we need $\log n$ bits of the record to encode the node's absolute position to cue the retrieval of the integer from the ILS. But the tag bit can be released during storing phase since we only need how many nodes are stored at the beginning of the list in total. Hence, we can use for instance the least significant $w-\log n$ bits of a record during practicing for any other purpose where  $w$ is the word length. It is immediate from this definition that a monotone bijective super hash function can be used during practicing. It should be noted that, this variant is suitable for sorting a {\em list} $S$ of $n$ {\em integers}, $S[0 \ldots n-1]$ where the problem is to sort the integers in ascending or descending order.

\subsection{Sorting Distinct Integers with Improved Technique}

Same idea that improves in-place associative sort can be used for sorting distinct integers, as well. This means that, if only the records ($w-1$ bits) of the nodes are clustered at the beginning of the list (short-term memory) retaining their relative order, there will be $n_d$ nodes dispersed in the list space, and $n_d$ records in the short-term memory ($S[0 \ldots n_d-1]$). Furthermore, there will be one-to-one correspondence between them.

With this introductory information, the contribution of this study is,

\begin{description}[leftmargin = 0pt]

\item[{\bf A practical algorithm}] that sorts $n$ {\em modifiable} {\em distinct} integers $S[0\ldots n-1]$ each in the range $[0,m-1]$ using $\mathcal{O}(1)$ extra space. If $\frac{m}{n} \le {w-1}$ the complexity of the algorithm is strictly $\mathcal{O}(n)$. Otherwise, it sorts the integers using $\mathcal{O}(1)$ extra space in $\mathcal{O}(n+\frac{m}{w-\log n})$ time for the worst, $\mathcal{O}(\frac{m}{w-\log n})$ time for the average (uniformly distributed integers) and $\mathcal{O}(n)$ time for the best case. Hence, the efficiency can be represented with $\frac{m}{n} \le c(w-1)$ where the constant $c > 1$ is determined by the other sorting algorithms. When improved modifiable distinct integer version is compared with read-only version, it has been observed that modifiable version is superior in every case. Hence, the only drawback of the algorithm when compared with the read-only version is that it is not suitable for {\em integer key sorting} where the problem is to sort the elements of the list according to their integer keys. 

\end{description}

\section{Definitions}\label{sec:pre}
The definition of {\em integer sorting} is: given a {\em list} $S$ of $n$ {\em integers}, $S[0 \ldots n-1]$, the problem is to sort the integers in ascending or descending order.

The notations used throughout the study are: 
\begin{enumerate} [label=({\roman{*}}), nosep]
\item Universe of integers is assumed $\mathbb{U} = [ 0 \ldots 2^{w}-1]$ where $w$ is the fixed word length.

\item Maximum and minimum integers of a list are, $\max (S) = \max(a \vert a \in S)$ and $\min (S) = \min(a \vert a \in S)$, respectively. Hence, range of the integers is, $m = \max (S) - \min (S) + 1$.

\item The notation $B \subset A$ is used to indicated that $B$ is a proper subset of $A$.

\item For two lists $S_{1}$ and $S_{2}$, $\max (S_{1}) < \min (S_{2})$ implies $S_{1} < S_{2}$.

\end{enumerate}

\begin{description}[leftmargin = 0pt]

\item[{\bf Universe of Integers.}] When an integer is first practiced, a node is created releasing $w$ bits of the integer free. One bit is used to tag the word as a node. Hence, it is reasonable to doubt that the tag bit limits the universe of integers because all the integers should be untagged and in the range $[0,2^{w-1}-1]$ before being practiced. But, we can,
\begin{enumerate}[label=(\roman{*}), itemindent = * , nosep]
\item partition $S$ into $2$ disjoint sublists $S_1 < 2^{w-1} \le S_2$ in $\mathcal{O}(n)$ time with well known in-place partitioning algorithms as well as stably with~\cite{Katajainen},
\item shift all the integers of $S_2$ by $-2^{w-1}$, sort $S_1$ and $S_2$ associatively and shift $S_2$ by $2^{w-1}$.
\end{enumerate}
There are other methods to overcome this problem. For instance, 
\begin{enumerate}[label=(\roman{*}), itemindent = * , nosep]
\item sort the sublist $S[0\ldots (n/ \log n)-1]$ using the optimal in-place merge sort~\cite{Salowe},
\item compress $S[0\ldots (n/ \log n)-1]$ by Lemma~1 of~\cite{Franceschini_1} generating $\Omega(n)$ free bits,
\item sort $S[(n/ \log n)\ldots n-1]$ associatively using $\Omega(n)$ free bits as tag bits,
\item uncompress $S[0\ldots (n/ \log n)-1]$ and merge the two sorted sublists in-place in linear time by~\cite{Salowe}.
\end{enumerate}

\item [{\bf Number of Integers.}] If practicing a distinct integer lets us to use $w-1$ bits to practice other occurrences of that integer, we have $w-1$ free bits by which we can count up to $2^{w-1}$ occurrences including the first integer that created the node. Hence, it is reasonable to doubt again that there is another restriction on the size of the lists, i.e., $n \le 2^{w-1}$. But a list can be divided into two parts in $\mathcal{O}(1)$ time and those parts can be merged in-place in linear time by~\cite{Salowe} after sorted associatively.

\end{description}

It should be noted that these restrictions are only valid for the variant proposed for modifiable integers. Hence, for the sake of simplicity, it will be assumed that $n \le 2^{w-1}$ and all the integers are in the range $[0,2^{w-1}-1]$ throughout the study.


\section{Sorting $n$ Distinct Modifiable Integers}\label{subsec:es_multiple}

In this section, the improved associative sorting technique for distinct modifiable integers will be introduced with its three basic steps: (i) practicing, (ii) storing and (iii) retrieval. 

Once a node is created for a particular integer when it is practiced, the redundancy due to the association between the integer and the node releases the word allocated for the integer in the physical memory except one bit which is used to tag the word as node of the ILS for interrogation. The released $w-1$ bits of a node become its record. Hence, we can use $w-1$ bits of a record during practicing for any other purpose. It is immediate from this definition that,
\begin{lem}\label{lem:sorting_seq_of_distinct_m}
Given $n$ distinct integers $S[0...n-1]$ each in the range $[u, v]$, all the $n_d$ integers in the range $[\delta,\delta+(w-1)n-1]$ with $\delta=\min(S)$ can be sorted associatively at the beginning of the list in $\mathcal{O}(n)$ time using only $\mathcal{O}(1)$ constant space.
\end{lem}

Given $n$ distinct integers $S[0...n-1]$ each in the range $[u, v]$, a monotone bijective {\em super} hash function can be constructed as a partial function that assigns each integer of $S_1 \subset S$ in the range $[\delta,\delta+(w-1)n-1]$ with $\delta=\min(S)$ to exactly one element in $j \in [0,n-1]$ and one element in $k \in [0,(w-1)-1]$. The simplest monotone bijective partial super hash function of this form is,
\begin{equation}\label{eqn:shf_1}
j = (S[i] - \delta) \, \idiv \, (w - 1)  \quad  \text{if} \quad S[i] - \delta < (w-1)n  
\end{equation}
\begin{equation}\label{eqn:shf_2}
k = (S[i] - \delta) \imod (w - 1)   \quad \text{if} \quad S[i] - \delta < (w-1)n 
\end{equation}
In this case, $w - 1$ integers may collide and mapped to the same node created at $j \in [0,n-1]$ (Eqn.~\ref{eqn:shf_1}) in the ILS. But we can use $w- 1$ free bits of a record of the node to encode which of $w-1$ distinct integers are mapped to the same node by setting the corresponding bit determined by $k$ (Eqn.~\ref{eqn:shf_2}). In other words, now the ILS is two dimensional over the list space where the first dimension along the list designates the node position and the second dimension along the bits of the node uniquely determines the integers which are mapped to the ILS through that node.

\begin{proof}
With this definition, the proof has three basic steps of associative sort: 

\begin{enumerate}[label=(\roman{*})]
\item Practice all the distinct integers of the interval $[\delta,\delta+(w-1)n-1]$ into $Im[0 \ldots n-1]$ over $S[0 \ldots n-1]$.

\item Store only the records ($w-1$ bits) of the nodes at the beginning of the list (short-term memory) retaining their relative order. Hence, a one-to-one correspondence is obtained with the stored records and the nodes (tagged words) of the list. 

\item Retrieve the sorted permutation of the practiced interval by searching the tagged words of the list backwards. When a tagged word (node) is found, retrieve the base of the integers from the ILS using the position of the node and the inverse of Eqn.\ref{eqn:shf_1}. Then, sequentially read the position of the bits that are equal to $1$ in the record which uniquely determines (with the inverse of Eqn.\ref{eqn:shf_2}) the integers mapped to the ILS through that node and expand them over the list backwards.

\end{enumerate}

\end{proof}

\subsection{Practicing Phase}\label{sec:counting_mul}

Details of practicing phase can be found in~\cite{ecetin, ecetin1, ecetin2}.

\begin{enumerate}[label=\bf{Algorithm \Alph{*}.}, ref=Algorithm \Alph{*}, leftmargin=0pt, itemindent=*, start=1] 
\item \label{algorithm:es_fgl_m} Practice all the distinct integers of the interval $[\delta,\delta+(w-1)n-1]$ by mapping them to the node determined by Eqn.~\ref{eqn:shf_1} in the ILS $Im[0...n-1]$ over $S[0...n-1]$. Once a integer is mapped to a node, set the integer's unique bit in the record determined by Eqn.~\ref{eqn:shf_2} which discriminates it from the others mapped to the same node. It is assumed that minimum of the list $\delta = \min{S}$ is known.
\end{enumerate}

\begin{enumerate}[label=\bf{A\arabic{*}.}, ref=A\arabic{*}, itemindent=*]
\item set $i = 0$;\label{algo5:item0}
\item if $S[i] < \delta$, then $S[i]$ is an idle integer of an interval that has already been sorted in the previous iterations (or recursions). Hence, increase $i$ and repeat this step;\label{algo5:item1}
\item if MSB of $S[i]$ is $1$, then $S[i]$ is a node. Hence, increase $i$ and goto step \ref{algo5:item1};\label{algo5:item2}
\item if $S[i] - \delta \ge (w-1)n$ then $S[i]$ is a integer of $S_2$ that is out of the practiced interval. Increase $n_d'$ that counts the number of integers of $S_2$, update $\delta'=min(\delta', S[i])$, increase $i$ and goto to step \ref{algo5:item1};\label{algo5:item3}
\item calculate $j$ and $k$ using Eqn.~\ref{eqn:shf_1} and \ref{eqn:shf_2}, respectively;\label{algo5:item4}
\item if MSB of $S[j]$ is $0$, then $S[i]$ is the first occurrence which will create the node at $S[j]$. Hence, move $S[j]$ to $S[i]$, clear $S[j]$ and set MSB and $k$th bit of $S[j]$ to $1$. If $j \le i$ increase $i$. Increase $n_d$ that counts the number of distinct integers and hence the nodes, and goto step \ref{algo5:item1}.\label{algo5:item5}
\item otherwise, a node has already been created at $S[j]$ by another occurrence of $S[i]$. Hence, set $k$th bit of $S[j]$ (without touching others) and increase $i$ and $n_c$ that counts number of total idle integers over all distinct integers, and goto step \ref{algo5:item1};
\end{enumerate}

\subsection{Storing Phase}\label{sec:partitioning_mul}

Practicing creates $n_d$ nodes and $n_c$ idle integers. This means $n_d$ integers of $S_1$ are mapped into the ILS creating nodes that are dispersed with relative order in $Im[0 \ldots n-1]$ over $S[0 \ldots n-1]$ depending on the statistical distribution of the integers. On the other hand, $n_c$ idle integers of $S_1$ are distributed disorderly together with $n_d'$ integers of $S_2$ in the list space.

In storing phase, the records are clustered in a systematic way to close the distance between them to a direction (beginning of the list) without altering their relative order with respect to each other. As long as the position of the nodes (tag bits) are not altered, the association between the ILS and the list space is retained as well as a one-to-one correspondence is attained between the records and the nodes. 

\begin{enumerate}[label=\bf{Algorithm \Alph{*}.}, ref=Algorithm \Alph{*}, leftmargin=0pt, itemindent=*, start=2] 
\item \label{algorithm:es_fgp_mul} Store the records of the practiced interval in the short term memory.  
\end{enumerate}
\begin{enumerate}[label=\bf{B\arabic{*}.}, ref=B\arabic{*}, itemindent=* , nosep]
\item initialize $i = 0$, $j = 0$, $k = n_d$;
\item if MSB of $S[i]$ is $0$, then $S[i]$ is either an idle integer or an integer of $S_2$ that is out of the practiced interval. Hence, increase $i$ and repeat this step; \label{algo2:item1}
\item otherwise, $S[i]$ is a node. Hence, swap least significant $w-1$ bits of $S[i]$ with least significant $w-1$ bits of $S[j]$. Increase $i$ and $j$ and decrease $k$. If $k = 0$ exit, otherwise goto step \ref{algo2:item1}; \label{algo2:item2}
\end{enumerate}

\subsection{Retrieval Phase}\label{sec:decoding_mul}

Storing clusters $n_d$ records of the nodes at $S[0 \ldots n_d-1]$. Hence, $S[0 \ldots n_d-1]$ can be though of as a short-term memory where the encoded information of the $n_d+n_c$ integers of the practiced interval is stored. 

In retrieval phase, the stored information is retrieved from the short term memory $S[0 \ldots n_d-1]$ to construct the sorted permutation of the practiced interval. The short term memory encodes $n_d+n_c$ integers of $S_1$ with $n_d$ permanent records. The stored information into a particular record is the unique bits of the integers (determined by Eqn.~\ref{eqn:shf_2}) practiced by the corresponding node. Hence, the nodes (tagged words) of the list have one-to-one correspondence with these $n_d$ records from left to right or vice versa. Hence, the base of the integers practiced by a node can be retrieved back to list space through the position of the node. It is important to note that, if the number of integers mapped to a node is $n_i$, then there are $n_i-1$ idle integers in the list. But the record itself represents an integer mapped into the ILS. Hence, it is immediate from this definition that the list can be searched from right to left backwards for the first node (tagged word) to retrieve the base of the integers practiced by that node (using the inverse of ~\ref{eqn:shf_1}) and the distinct integers can be calculated from their unique bits in the record $S[n_d-1]$ of the node in the short term memory $S[0 \ldots n_d-1]$ and expanded over the list backwards. Afterwards, the processed tag bit can be cleared and a new search to the right can be carried for the next node which will correspond to the next record $S[n_d-2]$ of the short-term memory. This can continue until all the integers are retrieved resulting in the sorted permutation of the practiced integers. 

It should be noted that, $n_c$ idle integers of $S_1$ and $n_d'$ integers of $S_2$ are distributed disorderly together at $S[n_d \ldots n-1]$. Hence, before proceeding, $n_c$ idle integers should be clustered at the beginning of $S[n_d \ldots n-1]$ because the practiced integers will be expanded over $S[0 \ldots n_d+n_c-1]$. This is a simple partitioning problem. However, the tag bits should be taken under care.

\begin{enumerate}[label=\bf{Algorithm \Alph{*}.}, ref=Algorithm \Alph{*}, leftmargin=0pt, itemindent=*, start=3] 
\item \label{algorithm:es_fgpp_mul} Partition $S[n_d \ldots n-1]$ to cluster $n_c$ idle integers to the beginning. 
\end{enumerate}
\begin{enumerate}[label=\bf{C\arabic{*}.}, ref=C\arabic{*}, itemindent=* , nosep]
\item initialize $i = n_d$, $j = n_d$, $k = n_c$;
\item Read $w-1$ bits of $S[i]$ into $s$. If $s - \delta >= n$, then $S[i]$ is an integer of $S_2$ that is out of the practiced interval. Hence, increase $i$ and repeat this step; \label{algo3:item1}
\item otherwise, $S[i]$ is an idle integer. Hence, swap least significant $w-1$ bits of $S[i]$ with least significant $w-1$ bits of $S[j]$. Increase $i$ and $j$ and decrease $k$. If $k = 0$ exit, otherwise goto step \ref{algo3:item1}; \label{algo3:item2}
\end{enumerate}

Afterwards, the retrieval phase can begin.
\begin{enumerate}[label=\bf{Algorithm \Alph{*}.}, ref=Algorithm \Alph{*}, leftmargin=0pt, itemindent=*, start=4] 
\item \label{algorithm:es_fgd_m}Process the list from right to left to find a node, retrieve the base of the integers mapped to that node from the ILS using inverse of Eqn.~\ref{eqn:shf_1}, sequentially read the position of the bits that are equal to $1$ from the record $S[n_d-1]$ which uniquely determines the corresponding integers mapped to that node and expand them over the list $S[0 \ldots n_d+n_c-1]$ right to left backwards. Then find the next node which corresponds to the record at $S[n_d-2]$. Continue until all the integers are retrieved and expanded over $S[0 \ldots n_d+n_c-1]$.
\end{enumerate}
\begin{enumerate}[label=\bf{D\arabic{*}.}, ref=D\arabic{*}, itemindent=*]
\item initialize $i = n_d-1$, $p = n_d + n_c$;
\item if MSB of $S[i]$ is $0$, then $S[i]$ is not a node, hence decrease $i$ and repeat this step;\label{algo7:item1_1}

\item initialize $k = w-2$ and calculate base of integers mapped to this node by $j = i(w-1)+\delta$ (inverse of Eqn.~\ref{eqn:shf_1}) using position $i$ of the node;\label{algo7:item1}
\item while $k \ge 0$; 
\begin{enumerate}[label=(\roman{*})]
\item if $k$th bit of $S[i]$ is $1$, then $S[p-1] = j + k$. Decrease $k$ and $p$ and repeat this step. 
\item otherwise, only decrease $k$ and goto step (i).
\end{enumerate}
\item clear MSB of $S[i]$, decrease $i$ and $p$. If $p=0$, then exit. Otherwise goto step \ref{algo7:item1_1};\label{algo7:item2}
\end{enumerate}

\begin{description}[leftmargin = 0pt]
\item [{\bf Sequential Version}] After retrieval phase, $n_d'$ integers of $S_2$ at $S[n_d+n_c \ldots n-1]$ can be sorted with the same algorithm using the minimum $\delta'$ found at step \ref{algo5:item3}. 

\begin{rem}
Improved associative sort technique is on-line in the sense that after each retrieval phase (\ref{algorithm:es_fgd_m}), $n_d+n_c$ integers are added to the sorted permutation at the beginning of the list and ready to be used.
\end{rem}

\begin{rem}
Recursive version is not possible as long as the tag bits are not clustered during storing phase.
\end{rem}

\end{description}

\begin{description}[leftmargin = 0pt] 
\item[{\bf Worst Case Complexity}] \ref{algorithm:es_fgl_m} to \ref{algorithm:es_fgd_m} are capable of sorting integers that satisfy $S[i] - \delta < (w-1)n$ in $\mathcal{O}(n)$ time. If we assume $m = \beta n$ with $\beta > w-1$, and there is only one integer available that satisfies $S[i] - \delta < (w-1)n$ in each iteration or recursion until the last, in any $j$th step, the only integer $s$ of $S_1$ that will be sorted satisfies,
\begin{equation}\label{eqn:distinct_m_wc}
s - \delta < j(w-1)n-(j-1)
\end{equation}
Eqn.~\ref{eqn:distinct_m_wc} implies that the last alone integer of $S$ satisfies,
\begin{equation}\label{eqn:distinct_m_wc_1}
s - \delta  < j(w-1)n-(j-1) \le \beta n
\end{equation}
from where can calculate $j$ by,
\begin{equation}
j \le \frac{\beta n-1}{(w-1)n - 1}
\end{equation}
In this case, the time complexity of the algorithm is 
\begin{equation}\label{eqn:distinct_m_wc_2}
\begin{split}
\mathcal{O}(n) + & \mathcal{O}(n-1) + \dotsc  + \mathcal{O}(n-j) = (j+1) \mathcal{O}(n) -\mathcal{O}(j^2) < (\frac{\beta}{w-1}+1) \mathcal{O}(n)
\end{split}
\end{equation}
Therefore, the time complexity of the algorithm in worst case is upper bounded by $\mathcal{O}(n+\frac{m}{w-1})$.


\item[{\bf Best Case}] If $n-1$ integers satisfy $S[i] - \delta < (w-1)n$, then these are sorted in $\mathcal{O}(n)$ time. In the next step, there is $n'=1$ integer left which implies sorting is finished. As a result, time complexity of the algorithm is lower bounded by $\Omega(n)$ in the best case.


\item[{\bf Average Case}] If we assume $m = \beta n (w-1)$ with $\beta> 1$, and the integers are uniformly distributed, then $\frac{n}{\beta}$ integers satisfy $S[i] - \delta < (w-1)n$. Therefore, the algorithm is capable of sorting $ \frac{n}{\beta}$ integers of the list in $\mathcal{O}(n)$ time at first step and $n'=n-\frac{n}{\beta}=\frac{n(\beta-1)}{\beta}$ integers will be left where $\frac{n'}{\beta}$ of them will be sorted in the next step. This will continue until all the integers are sorted. The complexity in this case is exactly equal to the complexity that we obtained for associative sorting of read-only distinct integers~\cite{ecetin1}. Hence, the time complexity of the sorting algorithm is upper bounded by $ \beta \mathcal{O}(n)$ or $\mathcal{O}(\frac{m}{w-1})$ for uniformly distributed lists.

\item[{\bf Practical Experience}] When improved modifiable distinct integer version is compared with read-only version, it has been observed that it is superior in every case.

\end{description}

\section{Conclusions}
\label{chap:summaryandconclusion}

In this study, the variant for modifiable distinct integers is improved and a practical algorithm is obtained that sorts $n$ distinct integers $S[0 \ldots n-1]$ each in the range $[0, m - 1]$ in exactly $\mathcal{O}(n)$ time if $m < (w-1) n$. Otherwise, it sort in $\mathcal{O}(n+\frac{m}{w-1})$ time for the worst,  $\mathcal{O}(\frac{m}{w-1})$ time for the average (uniformly distributed keys) and $\mathcal{O}(n)$ time for the best case using only $\mathcal{O}(1)$ extra space.

\end{document}